\documentstyle[12pt]{article}
\begin{document}

\title{\bf{Interacting entropy-corrected new agegraphic dark energy in non-flat universe}}
\author{K. Karami$^{1,2}$\thanks{E-mail: KKarami@uok.ac.ir} , A. Sorouri$^{1}$\\
$^{1}$\small{Department of Physics, University of Kurdistan,
Pasdaran St., Sanandaj, Iran}\\$^{2}$\small{Research Institute for
Astronomy $\&$ Astrophysics of Maragha (RIAAM), Maragha, Iran}}
\maketitle

\begin{abstract}
Here we consider the entropy-corrected version of the new
agegraphic dark energy model in the non-flat FRW universe. We
derive the exact differential equation that determines the
evolution of the entropy-corrected new agegraphic dark energy
density parameter in the presence of interaction with dark matter.
We also obtain the equation of state and deceleration parameters
and present a necessary condition for the selected model to cross
the phantom divide. Moreover, we reconstruct the potential and the
dynamics of the phantom scalar field according to the evolutionary
behavior of the interacting entropy-corrected new agegraphic
model.
\end{abstract}
\clearpage
\section{Introduction}
Type Ia supernovae observational data suggest that the universe is
dominated by two dark components: dark matter and dark energy
\cite{Riess}. Dark matter (DM), a matter without pressure, is
mainly used to explain galactic curves and large-scale structure
formation, while dark energy (DE), an exotic energy with negative
pressure, is used to explain the present cosmic accelerating
expansion. However, the nature of DE is still unknown, and people
have proposed some candidates to describe it (for review see
\cite{Karami1,Copeland} and references therein).

The holographic DE (HDE) is one of interesting DE candidates which
was proposed based on the holographic principle \cite{Horava}.
According to the holographic principle, the number of degrees of
freedom in a bounded system should be finite and has relations
with the area of its boundary \cite{Hooft}. By applying the
holographic principle to cosmology, one can obtain the upper bound
of the entropy contained in the Universe \cite{Fischler}.
Following this line, Li \cite{Li} argued that for a system with
size $L$ and UV cut-off $\Lambda$ without decaying into a black
hole, it is required that the total energy in a region of size $L$
should not exceed the mass of a black hole of the same size, thus
$L^3\rho_{\Lambda}\leq LM_P^2$, where $\rho_{\Lambda}$ is the
quantum zero-point energy density caused by UV cut-off $\Lambda$
and $M_P$ is the reduced Planck mass $M_P^{-2}=8\pi G$. The
largest $L$ allowed is the one saturating this inequality, thus
$\rho_{\Lambda}=3c^2M_P^2L^{-2}$, where $c$ is a numerical
constant. The HDE models have been studied widely in the
literature \cite{Enqvist,Elizalde2,Guberina1,Guberina2}.
Obviously, in the derivation of HDE, the black hole entropy
$S_{\rm BH}$ plays an important role. As is well known, usually,
$S_{\rm BH} = A/(4G)$, where $A\sim L^2$ is the area of horizon.
However, in the literature, this entropy-area relation can be
modified to \cite{modak}
\begin{equation}
S_{\rm
BH}=\frac{A}{4G}+\tilde{\alpha}\ln{\frac{A}{4G}}+\tilde{\beta},\label{MEAR}
\end{equation}
where $\tilde{\alpha}$ and $\tilde{\beta}$ are dimensionless
constants of order unity. These corrections can appear in the
black hole entropy in loop quantum gravity (LQG) \cite{HW}. They
can also be due to thermal equilibrium fluctuation, quantum
fluctuation, or mass and charge fluctuations (for a good review
see \cite{HW} and references therein). Using the corrected
entropy-area relation (\ref{MEAR}), the energy density of the
entropy-corrected HDE (ECHDE) can be obtained as \cite{HW}
\begin{equation}
\rho_{\Lambda}=3c^2M_P^2L^{-2}+\alpha L^{-4}\ln(M_P^2L^{2})+\beta
L^{-4},\label{ECHDE}
\end{equation}
where $\alpha$ and $\beta$ are dimensionless constants of order
unity.

Recently, the original agegraphic dark energy (OADE) and new
agegraphic dark energy (NADE) models were proposed by Cai
\cite{Cai} and Wei $\&$ Cai \cite {Wei1}, respectively. These
models are based on the uncertainty relation of quantum mechanics
as well as the gravitational effect in general relativity.
Following the line of quantum fluctuations of spacetime,
Karolyhazy et al. \cite{Kar1} argued that the distance $t$ in
Minkowski spacetime cannot be known to a better accuracy than
$\delta{t}\sim t_{P}^{2/3}t^{1/3}$ where $t_P$ is the reduced
Planck time. Based on Karolyhazy relation, Maziashvili \cite{Maz}
discussed that the energy density of metric fluctuations of the
Minkowski spacetime is given by
\begin{equation}
\rho_{\Lambda}\sim \frac{1}{t_P^2t^2}\sim
\frac{M_P^2}{t^2}.\label{rhoMaz}
\end{equation}
Based on Karolyhazy relation \cite{Kar1} and Maziashvili arguments
\cite{Maz}, Cai proposed the OADE model to explain the accelerated
expansion of the universe \cite{Cai}. The ADE models assume that
the observed DE comes from the spacetime and matter field
fluctuations in the universe \cite{Cai}. The OADE model had some
difficulties. In particular, it cannot justify the
matter-dominated era \cite{Cai}. This motivated Wei and Cai
\cite{Wei1} to propose the NADE model, while the time scale is
chosen to be the conformal time instead of the age of the
universe. The evolution behavior of the NADE is very different
from that of the OADE. Instead the evolution behavior of the NADE
is similar to that of the HDE. But some essential differences
exist between them. In particular, the NADE model is free of the
drawback concerning causality problem which exists in the HDE
model. The ADE models have arisen a lot of enthusiasm recently and
have examined and studied in ample detail by
\cite{Wei2,Kim,Kim11,Wei3,Sheykhi}.

Here our aim is to investigate the entropy-corrected version of
the interacting NADE model in the non-flat universe. This paper is
organized as follows. In Section 2, we give a brief review on the
NADE and the entropy-corrected NADE (ECNADE) models. In Section 3,
we study the ECNADE in a FRW universe with spacial curvature and
in the presence of interaction between DE and DM. In Section 4, we
reconstruct the potential and the dynamics of the phantom scalar
field according to the evolutionary behavior of the interacting
ECNADE model. Section 5 is devoted to conclusions.
\section{A brief review on the NADE and ECNADE models}

From Eq. (\ref{rhoMaz}), the energy density of the NADE is given
by \cite{Wei1}
\begin{equation}
\rho_{\Lambda}=\frac{3{n}^2M_P^2}{\eta^2},\label{NADE}
\end{equation}
where the numerical factor 3$n^2$ is introduced to parameterize
some uncertainties, such as the species of quantum fields in the
universe, the effect of curved spacetime (since the energy density
is derived for Minkowski spacetime), and so on. It was found that
the coincidence problem could be solved naturally in the NADE
model provided that the single model parameter $n$ is of order
unity \cite{Wei3}. Also $\eta$ is conformal time of the FRW
universe, and given by
\begin{equation}
\eta=\int\frac{{\rm d}t}{a}=\int\frac{{\rm d}a}{Ha^2}.\label{eta}
\end{equation}
With the help of quantum corrections to the entropy-area relation
(\ref{MEAR}) in the setup of LQG, the energy density of the ECNADE
is given by \cite{HW}
\begin{eqnarray}
\rho_{\Lambda} = \frac{3n^2{M_P^2}}{\eta^2} +
\frac{\alpha}{{\eta}^4}\ln{({M_P^2}{\eta}^2)} +
\frac{\beta}{\eta^4},\label{density-nade}
\end{eqnarray}
which closely mimics to that of ECHDE density (\ref{ECHDE}) and
$L$ is replaced with the conformal time $\eta$. Here $\alpha$ and
$\beta$ are dimensionless constants of order unity. In the special
case $\alpha=\beta=0$, the above equation yields the well-known
NADE density (\ref{NADE}).
\section{Interacting ECNADE and DM in non-flat universe}

We consider the Friedmann-Robertson-Walker (FRW) metric for the
non-flat universe as
\begin{equation}
{\rm d}s^2=-{\rm d}t^2+a^2(t)\left(\frac{{\rm
d}r^2}{1-kr^2}+r^2{\rm d}\Omega^2\right),\label{metric}
\end{equation}
where $k=0,1,-1$ represent a flat, closed and open FRW universe,
respectively. Observational evidences support the existence of a
closed universe with a small positive curvature ($\Omega_{k}\sim
0.02$) \cite{Bennett}. Besides, as usually believed, an early
inflation era leads to a flat universe. This is not a necessary
consequence if the number of e-foldings is not very large
\cite{Huang}. It is still possible that there is a contribution to
the Friedmann equation from the spatial curvature when studying
the late universe, though much smaller than other energy
components according to observations.

For the non-flat FRW universe containing the DE and DM, the first
Friedmann equation has the following form
\begin{equation}
{\textsl{H}}^2+\frac{k}{a^2}=\frac{1}{3M_P^2}~
(\rho_{\Lambda}+\rho_{\rm m}),\label{eqfr}
\end{equation}
where $\rho_{\Lambda}$ and $\rho_{\rm m}$ are the energy density
of DE and DM, respectively. Let us define the dimensionless energy
densities as
\begin{equation}
\Omega_{\rm m}=\frac{\rho_{\rm m}}{\rho_{\rm cr}}=\frac{\rho_{\rm
m}}{3M_P^2H^2},~~~~~~\Omega_{\rm
\Lambda}=\frac{\rho_{\Lambda}}{\rho_{\rm
cr}}=\frac{\rho_{\Lambda}}{3M_P^2H^2},~~~~~~\Omega_{k}=\frac{k}{a^2H^2},
\label{eqomega}
\end{equation}
then, the first Friedmann equation yields
\begin{equation}
\Omega_{\rm m}+\Omega_{\Lambda}=1+\Omega_{k}.\label{eq10}
\end{equation}

We consider a universe containing an interacting ECNADE density
$\rho_{\Lambda}$ and the cold dark matter (CDM), with $\omega_{\rm
m}=0$. The energy equations for ECNADE and CDM are
\begin{equation}
\dot{\rho}_{\Lambda}+3H(1+\omega_{\Lambda})\rho_{\Lambda}=-Q,\label{eqpol}
\end{equation}
\begin{equation}
\dot{\rho}_{\rm m}+3H\rho_{\rm m}=Q,\label{eqCDM}
\end{equation}
where following \cite{Kim06}, we choose $Q=\Gamma\rho_{\Lambda}$
as an interaction term and
$\Gamma=3b^2H(\frac{1+\Omega_{k}}{\Omega_{\Lambda}})$ is the decay
rate of the ECNADE component into CDM with a coupling constant
$b^2$. This expression for the interaction term $Q$ was first
introduced in the study of the suitable coupling between a
quintessence scalar field and a pressureless CDM field
\cite{Amendola1}. Although at this point the interaction may look
purely phenomenological but different Lagrangians have been
proposed in support of it \cite{Tsujikawa}. Tsujikawa and Sami
\cite{Tsujikawa} investigated the cosmological scaling solutions
in a general cosmological background $H^2\propto
(\rho_{\Lambda}+\rho_{\rm m})^\varepsilon$ including general
relativity (GR), Randall-Sundrum (RS) braneworld and Gauss-Bonnet
(GB) braneworld. The GR, RS and GB cases correspond to
$\varepsilon = 1$, $\varepsilon = 2$ and $\varepsilon = 2/3$,
respectively. The condition for the existence of scaling solutions
restricts the form of the scalar field Lagrangian (or the pressure
density) to be
$p=X^{1/\varepsilon}g(Xe^{\varepsilon\lambda\phi})$, where $\phi$
is a scalar field with $X$ defined as
$X=-g^{\mu\nu}\partial_{\mu}\phi\partial_{\nu}\phi/2$ and $g$ is
an arbitrary function. Also $\lambda\propto Q$, where $Q$ is the
coupling term due to the interaction between the scalar filed and
the matter. Tsujikawa and Sami \cite{Tsujikawa} showed that in the
absence of the coupling $Q$ between a scalar field and a perfect
barotropic fluid, it is not possible to get an acceleration of the
universe since the energy density of the field $\phi$ decreases in
proportional to that of the background fluid for scaling
solutions. However the presence of the coupling $Q$ allows to have
an accelerated expansion. It should be emphasized that this
phenomenological description has proven viable when contrasted
with observations, i.e., SNIa, CMB, large scale structure, $H(z)$,
and age constraints \cite{Wang8}, and recently in galaxy clusters
\cite{Bertolami8}. The choice of the interaction between both
components was to get a scaling solution to the coincidence
problem such that the universe approaches a stationary stage in
which the ratio of DE and DM becomes a constant \cite{Hu}. The
dynamics of interacting DE models with different $Q$-classes have
been studied in ample detail by \cite{Amendola}.

From definition $\rho_{\Lambda}=3M_P^2H^2\Omega_{\Lambda}$, we get
\begin{eqnarray}
\Omega_{\Lambda} =
\frac{n^2}{H^2\eta^2}\gamma_{n},\label{density-nade-omega}
\end{eqnarray}
where
\begin{eqnarray}
\label{gamma-parameter1} \gamma_n = 1 +
\frac{1}{3n^2{M_P^2}\eta^2}\Big[\alpha\ln{({M_P^2}{\eta}^2)}
+\beta\Big].
\end{eqnarray}
Taking derivative of Eq. (\ref{density-nade-omega}) with respect
to $x=\ln a$, using $\dot{\eta}=1/a$ and
${\Omega^{\prime}_{\Lambda}}=\dot{\Omega_{\Lambda}}/H$ where prime
denotes the derivative with respect to $x$, one can obtain the
equation of motion for $\Omega_{\Lambda}$ as
\begin{eqnarray}
\label{omegaD-eq-motion1} {\Omega^{\prime}_{\Lambda}} =
-2\Omega_{\Lambda}
\Big[\frac{\dot{H}}{H^2}+\frac{1}{na\gamma_n}\Big({\frac{\Omega_{\Lambda}}{\gamma_n}}\Big)^{1/2}\Big(2\gamma_n
- 1-\frac{\alpha
H^2}{3{M^{2}_P}n^4}\frac{\Omega_{\Lambda}}{\gamma_n}\Big)\Big].
\end{eqnarray}
Taking time derivative of the first Friedmann equation
(\ref{eqfr}) and using Eqs. (\ref{density-nade}), (\ref{eqomega}),
(\ref{eq10}), (\ref{eqCDM}), (\ref{density-nade-omega}),
(\ref{gamma-parameter1})
 one can get
\begin{eqnarray}
\label{H-dot-to-H2} \frac{\dot{H}}{H^2} = -\frac{3}{2}(1 -
\Omega_{\Lambda}) +\frac{3b^2}{2}(1 + \Omega_k)-
\frac{\Omega_k}{2}
-\frac{1}{na}\Big({\frac{\Omega_{\Lambda}}{\gamma_n}}\Big)^{3/2}\Big(2\gamma_n
- 1-\frac{\alpha
H^2}{3{M^{2}_P}n^4}\frac{\Omega_{\Lambda}}{\gamma_n}\Big).
\end{eqnarray}
Substituting this into Eq. (\ref{omegaD-eq-motion1}), one obtains
\begin{eqnarray}
\label{omegaD-eq-motion2} {\Omega^{\prime}_{\Lambda}} =
\Omega_{\Lambda} \Big[3(1 - \Omega_{\Lambda})  - 3b^2(1 +
\Omega_k) +
\Omega_k+\frac{2}{na}\Big({\frac{\Omega_{\Lambda}}{\gamma_n}}\Big)^{1/2}\Big(2\gamma_n
-1-\frac{\alpha
H^2}{3{M^{2}_P}n^4}\frac{\Omega_{\Lambda}}{\gamma_n}\Big)\Big(\frac{\Omega_{\Lambda}-1}{\gamma_n}\Big)\Big].~~~
\end{eqnarray}
Taking time derivative of Eq. (\ref{density-nade}), using
$\dot{\eta}=1/a$ and substituting the result in Eq. (\ref{eqpol})
yields the equation of state (EoS) parameter of the interacting
ECNADE as
\begin{eqnarray}
\label{state-parameter} w_{\Lambda} = -1 - b^2\Big(\frac{1 +
\Omega_k}{\Omega_{\Lambda}}\Big)+
\frac{2}{3na\gamma_n}\Big({\frac{\Omega_{\Lambda}}{\gamma_n}}\Big)^{1/2}\Big(2\gamma_n
- 1-\frac{\alpha
H^2}{3{M^{2}_P}n^4}\frac{\Omega_{\Lambda}}{\gamma_n}\Big),
\end{eqnarray}
which shows that the interacting ECNADE can cross the phantom
divide, i.e. $\omega_{\Lambda}<-1$, when
\begin{equation}
3nab^2(1+\Omega_k)>2\Big(\frac{\Omega_{\Lambda}}{\gamma_n}\Big)^{3/2}\Big(2\gamma_n
- 1-\frac{\alpha
H^2}{3{M^{2}_P}n^4}\frac{\Omega_{\Lambda}}{\gamma_n}\Big).
\end{equation}
During the inflation era where $H=$ constant, we have $a=e^{Ht}$
and from Eq. (\ref{eta}) we get $\eta=-1/Ha$. Since the last two
terms in Eq. (\ref{density-nade}) can be comparable to the first
term only when $\eta$ is very small, the corrections make sense
only at the late stage of the inflationary expansion of the
universe. During the cosmological inflation in the early universe,
ECNADE reduces to the NADE model. The EoS parameter of the
interacting ECNADE during the inflation era will be
\begin{eqnarray}
\label{state-parameter1} w_{\Lambda} = -1 - b^2\Big(\frac{1 +
\Omega_k}{\Omega_{\Lambda}}\Big)+
\frac{2e^{-Ht}}{3n\gamma_n}\Big({\frac{\Omega_{\Lambda}}{\gamma_n}}\Big)^{1/2}\Big(2\gamma_n
- 1-\frac{\alpha H^2e^{2Ht}}{3{M^{2}_P}n^2}\Big),
\end{eqnarray}
with
\begin{eqnarray}
\label{gamma-parameter-2} \gamma_n = 1 +
\frac{H^2e^{2Ht}}{3n^2{M_P^2}}\Big[\alpha\ln{\Big(\frac{M_P^2}{H^2e^{2Ht}}\Big)}
+\beta\Big].
\end{eqnarray}
The deceleration parameter is given by
\begin{equation}
q=-\Big(1+\frac{\dot{H}}{H^2}\Big).\label{q1}
\end{equation}
Putting Eq. (\ref{H-dot-to-H2}) in the above relation reduces to
\begin{eqnarray}
\label{deceleration} q = - \frac{3}{2}\Omega_{\Lambda} +
\frac{1}{2}(1 - 3b^2)(1+\Omega_k)+
\frac{1}{na}\Big({\frac{\Omega_{\Lambda}}{\gamma_n}}\Big)^{3/2}\Big(2\gamma_n
- 1-\frac{\alpha
H^2}{3{M^{2}_P}n^4}\frac{\Omega_{\Lambda}}{\gamma_n}\Big).
\end{eqnarray}
Note that if we set $\alpha=\beta=0$ then from Eq.
(\ref{gamma-parameter1}) $\gamma_n=1$. Therefore Eqs.
(\ref{state-parameter}) and (\ref{deceleration}) reduce to
\begin{eqnarray}
\omega_{\Lambda}=-1-b^2\Big(\frac{1+\Omega_{k}}{\Omega_{\Lambda}}\Big)+\frac{2\Omega_{\Lambda}^{1/2}}{3{
n}a},\label{w}
\end{eqnarray}
\begin{equation}
q=-\frac{3}{2}\Omega_{\Lambda}+\frac{1}{2}(1-3b^2)(1+\Omega_{k})+\frac{\Omega_{\Lambda}^{3/2}}{na},\label{q3}
\end{equation}
which are the EoS and deceleration parameters of the interacting
NADE with CDM in the non-flat universe \cite{Sheykhi, Karami5}.

Following \cite{Cai}, the OADE density is given by
\begin{equation}
\rho_{\Lambda}=\frac{3{n}^2M_P^2}{T^2},\label{OADE}
\end{equation}
where $T$ is the age of the universe and given by
\begin{equation}
T=\int{\rm d}t=\int\frac{{\rm d}a}{Ha}.
\end{equation}
Note that appearing the age of the universe $T$ in the energy
density of the OADE model causes some difficulties. In particular
it fails to describe the matter-dominated epoch properly
\cite{Cai}. Similar to the density of ECNADE (\ref{density-nade}),
the density of entropy-corrected OADE (ECOADE) can be written as
\begin{eqnarray}
\rho_{\Lambda} = \frac{3n^2{M_P^2}}{T^2} +
\frac{\alpha}{{T}^4}\ln{({M_P^2}{T}^2)} +
\frac{\beta}{T^4}.\label{density-OADE}
\end{eqnarray}
To obtain the evolution of the density parameter, the EoS and
deceleration parameters of the interacting ECOADE in a non-flat
FRW universe, one doesn't need to repeat the calculations. The
only necessary changes are that one must replace $\eta$ with $T$
in Eq. (\ref{gamma-parameter1}) and put $a=1$ in Eqs.
(\ref{omegaD-eq-motion2}), (\ref{state-parameter}) and
(\ref{deceleration}).

\section{Entropy-corrected new agegraphic phantom model}
Here we suggest a correspondence between the interacting ECNADE
model with the phantom scalar field model. The phantom scalar
field model is often regarded as an effective description of an
underlying theory of DE \cite{Wu}. Recent observational data
indicates that the EoS parameter $\omega_{\Lambda}$ lies in a
narrow strip around $\omega_{\Lambda} = -1$ and is quite
consistent with being below this value. The region where the EoS
is less than $-1$ is typically referred to as a being due to some
form of phantom (ghost) DE \cite{Copeland}. However, although
fundamental theories such as string/M theory do provide this
scalar field, they do not uniquely predict its potential $V(\phi)$
\cite{Wu,Ali}. Therefore it becomes meaningful to reconstruct
$V(\phi)$ from some DE models possessing some significant features
of the LQG theory, such as ECHDE and ECNADE models.

The energy density and pressure of the phantom scalar field $\phi$
are as follows \cite{Copeland}
\begin{equation}
\rho_{{\rm ph}}=-\frac{1}{2}\dot{\phi}^{2}+V(\phi),\label{rhoq}
\end{equation}
\begin{equation}
p_{{\rm ph}}=-\frac{1}{2}\dot{\phi}^{2}-V(\phi).
\end{equation}
The EoS parameter for the phantom scalar field is given by
\begin{equation}
\omega_{{\rm ph}}=\frac{p_{{\rm ph}}}{\rho_{{\rm
ph}}}=\frac{\dot{\phi}^{2}+2V(\phi)}{\dot{\phi}^{2}-2V(\phi)}.\label{wq}
\end{equation}
Here like \cite{Karami2}, we establish the correspondence between
the interacting ECNADE scenario and the phantom DE model, then
equating Eq. (\ref{wq}) with the EoS parameter of interacting
ECNADE (\ref{state-parameter}), $\omega_{\rm
ph}=\omega_{\Lambda}$, and also equating Eq. (\ref{rhoq}) with
(\ref{density-nade}), $\rho_{\rm ph}=\rho_{\Lambda}$, we have
\begin{equation}
\dot{\phi}^2=-(1+\omega_{\Lambda})\rho_{\Lambda},\label{phidot2-2}
\end{equation}
\begin{equation}
V(\phi)=\frac{1}{2}(1-\omega_{\Lambda})\rho_{\Lambda}.\label{Vphi-2}
\end{equation}
Substituting Eqs. (\ref{density-nade}) and (\ref{state-parameter})
into Eqs. (\ref{phidot2-2}) and (\ref{Vphi-2}), one can obtain the
kinetic energy term and the phantom potential energy as follows
\begin{eqnarray}
\dot{\phi}^2=3M_{P}^2H^2\Big[b^2(1 +
\Omega_k)-\frac{2}{3na}\Big(\frac{\Omega_{\Lambda}}{\gamma_n}\Big)^{3/2}\Big(2\gamma_n
- 1-\frac{\alpha
H^2}{3{M^{2}_P}n^4}\frac{\Omega_{\Lambda}}{\gamma_n}\Big)\Big],\label{phi2q}
\end{eqnarray}
\begin{eqnarray}
V(\phi)=\frac{3}{2}M_{P}^2H^2\Big[2\Omega_{\Lambda}+b^2(1 +
\Omega_k)-\frac{2}{3na}\Big(\frac{\Omega_{\Lambda}}{\gamma_n}\Big)^{3/2}\Big(2\gamma_n
- 1-\frac{\alpha
H^2}{3{M^{2}_P}n^4}\frac{\Omega_{\Lambda}}{\gamma_n}\Big)\Big].\label{vphiq}
\end{eqnarray}
From Eq. (\ref{phi2q}) and using $\dot{\phi}=\phi'H$, one can
obtain the evolutionary form of the phantom scalar field as
\begin{eqnarray}
\phi(a)-\phi(a_0)=\sqrt{3}M_{P}\int_{\ln a_0}^{\ln a}\Big[b^2(1 +
\Omega_k)-\frac{2}{3na}\Big(\frac{\Omega_{\Lambda}}{\gamma_n}\Big)^{3/2}\Big(2\gamma_n
- 1-\frac{\alpha
H^2}{3{M^{2}_P}n^4}\frac{\Omega_{\Lambda}}{\gamma_n}\Big)\Big]^{1/2}{\rm
d}x,\label{phi3q}
\end{eqnarray}
where $a_0$ is the scale factor at the present time.

\section{Conclusions}

Here we considered the entropy-corrected version of the NADE model
which is in interaction with CDM in the non-flat FRW universe.
However, some experimental data have implied that our universe is
not a perfectly flat universe and that it possesses a small
positive curvature ($\Omega_{k}\sim0.02$) \cite{Bennett}. Although
it is believed that our universe is flat, a contribution to the
Friedmann equation from spatial curvature is still possible if the
number of e-foldings is not very large \cite{Huang}. The ADE
models proposed to explain the accelerated expansion of the
universe, based on the uncertainty relation of quantum mechanics
as well as the gravitational effect in general relativity
\cite{Cai,Wei1}. We considered the logarithmic correction term to
the energy density of NADE model. The addition of correction terms
to the energy density of NADE is motivated from the LQG which is
one of the promising theories of quantum gravity. Using this
modified energy density, we derived the exact differential
equation that determines the evolution of the ECNADE density
parameter. We also obtained the EoS and deceleration parameters
for the interacting ECNADE and present a necessary condition for
the present model to cross the phantom divide. Moreover, we
established a correspondence between the interacting ECNADE
density and the phantom scalar field model of DE. We adopted the
viewpoint that the scalar field models of DE are effective
theories of an underlying theory of DE. Thus, we should be capable
of using the scalar field model to mimic the evolving behavior of
the interacting ECNADE and reconstructing this scalar field model.
We reconstructed the potential and the dynamics of the phantom
scalar field, which describe accelerated expansion of the
universe, according to the evolutionary behavior of the
interacting ECNADE model.
\\
\\
\noindent{{\bf Acknowledgements}}\\
The authors thank the reviewers for very valuable comments. This
work has been supported financially by Research Institute for
Astronomy $\&$ Astrophysics of Maragha (RIAAM), Maragha, Iran.


\end{document}